\documentstyle[11pt,aaspp4]{article}
\def\deg{\arcdeg}

\begin{document}

\title{Serendipitous Discovery of a BAL QSO at z = 2.169\footnotemark[1]}

\footnotetext[1]{Based on observations made with the NASA/ESA Hubble
Space Telescope, obtained from the data archive at the Space Telescope
Science Institute, which is operated by the Association of Universities
for Research in Astronomy, Inc., under NASA contract NAS 5-26555.}

\author {Gabriela Canalizo\altaffilmark{2}, Alan Stockton\altaffilmark{2},
and Katherine C. Roth\altaffilmark{3}}
\affil{Institute for Astronomy, University of Hawaii, 2680 Woodlawn
 Drive, Honolulu, HI 96822}

\altaffiltext{2}{Visiting Astronomer, W.M. Keck Observatory, jointly operated
by the California Institute of Technology and the University of California.}

\altaffiltext{3}{Hubble Fellow.}

\begin{abstract}
We report the serendipitous discovery of a BAL QSO at z = 2.169, located
41\arcsec\ southwest of 3C\,48.   We present Keck LRIS spectroscopy covering 
rest frame 1500 \AA\ to 2300 \AA .  The C\,IV BAL has three components and 
it extends to outflow velocities of at least 12,000 km s$^{-1}$. 

The BAL QSO has an intervening low-ionization metal-line absorption system 
at z = 1.667 which is likely to be a damped Ly$\alpha$ absorber.  HST images 
show extended luminous material around the QSO, which could be either the 
host galaxy or the intervening system.
\end{abstract}


\section{Introduction}
Broad Absorption Line Quasi Stellar Objects (BAL QSOs) form a rare class
comprising  $\sim12$\% of the QSO population at moderate to high redshift
(Foltz et al.\ 1990).  The standard view, based largely on estimates of
upper limits to resonant scattering by the BAL clouds, is that such
clouds have a small covering factor as seen from the QSO nucleus, 
implying that essentially all radio-quiet QSOs would be classified
as BAL QSOs if observed from the proper angle.  However,
BAL QSOs appear to be found more frequently among
those with enhanced far-IR emission, strong \ion{Fe}{2} emission, and/or
weak or absent [\ion{O}{3}] narrow emission lines (Low et al.\ 1989; 
Boroson \& Meyers 1992; Turnshek et al.\ 1997).
While the strong \ion{Fe}{2} emission might be due to an orientation
effect, it seems unlikely that the other properties mentioned would
have a strong dependence on viewing angle.

One interpretation of the BAL phenomenon running counter to the standard
unified view is that these QSOs are very young 
and are ejecting their gaseous envelopes at very high velocities
following the initial turn-on of the AGN (Hazard et al. 1984). 
In particular, the high incidence of low-ionization 
BAL QSOs in IRAS-selected, luminous QSOs, along with the reddening from 
dust in these objects, suggests that this subclass, at least,
may comprise transition objects between ultraluminous infrared galaxies and 
the classical QSO population (L\'{\i}pari 1994, Voit et al 1993, Egami et al. 
1996).

In principle, characterization of the host galaxies of sufficiently
large samples of normal radio-quiet QSOs and BAL QSOs might tell us
whether the BAL phenomenon is mostly due to orientation or mostly due
to intrinsic properties.
Here we report the chance discovery of a BAL QSO at moderately 
high redshift, for which HST imaging already exists.  We found the QSO in 
the field of the quasar 3C\,48 as part of
a Keck spectroscopic program studying the host galaxies of low redshift QSOs
(Canalizo \& Stockton 1997, 1998).  
The only previous mention of this object in the astronomical literature is
its identification as a ``star'' by Kirhakos et al.\ (1995).
It is because of its proximity to 3C\,48
that it is present on HST WFPC archival images, and, as we shall
describe, it does show some extended emission.

\section{Observations and Data Reduction}

Spectroscopic observations of the host galaxy of 3C\,48 were carried out
on 1996 October 13 
with the Low-Resolution Imaging Spectrometer (LRIS) on the Keck II telescope. 
We used a 600 groove mm$^{-1}$ grating blazed at 5000\,\AA \ to obtain a 
dispersion of 1.24\,\AA \,pixel$^{-1}$ and a useful wavelength range of 
4600--7200\,\AA.   One of our slit positions happened to include a stellar
object to the south of 3C\,48.
 The slit (1\arcsec\ wide, projecting to $\sim$5 pixels on 
the Tektronix 2048$\times$2048 CCD)  was centered $\sim$1\farcs5 E of
3C\,48, with a PA of 14\deg, as indicated on Fig. 1. The spectrum of
what we will refer to as the BAL QSO 0134+3253 fell 41\arcsec\ below that of 
the host galaxy of 3C\,48.  The total exposure time was 60 minutes.

The spectra were reduced with IRAF, using standard reduction procedures.
After dark, bias and flat field correction, the images were
background subtracted to remove sky lines.   Wavelength calibrations
were made using the least-mean-squares fit to identified lines in a
comparison spectrum.  
The spectra were flux calibrated using spectrophotometric standards from
Massey {\em et al.} (1988).  The distortions in the spatial coordinate
were removed with the {\em apextract} routines.

WFPC2 images of the 3C\,48 field were obtained from the HST data archive.
The images used in this analysis included two 1400 s exposures in the F555W
filter and two 1700 s exposures in the F814W filter.  In each case, 3C\,48
was centered on the PC1 detector and the BAL QSO fell near pixel coordinates
(308, 244) in the WFC2 detector.
Most cosmic rays were removed by subtracting one of the frames (``image 2'')
from the other frame obtained with the same filter (``image 1''), then 
thresholding the difference at a
3$\sigma$ level, setting points above this threshold to the median of
the difference image.  Pixels near the position of the peak of the QSO
were excluded from this process.  The corrected difference image was then
added back to image 2, giving a corrected version of image 1.  The few
cosmic rays within the relevant region that escaped this process were 
removed manually with the IRAF task {\it imedit}.  The procedure was repeated,
interchanging the images, to correct image 2, and the two corrected images
for each filter were averaged.  

In order to study extended luminous material around the QSO, we subtracted
a model point-spread function (PSF).  This PSF was based on a star of similar
brightness to the QSO, located 15\arcsec\ away.  While the WFPC2 shows 
significant field variations in the PSF, at the detection level of the QSO
features of the PSF beyond a radius of 0\farcs4 are insignificant compared
with the noise.  We found we could achieve a good match simply with a
slight adjustment of the ellipticity of the PSF model.

Finally, the field including 3C\,48 and the BAL QSO was imaged with a notched
$H + K'$ filter on 1997 October 2 with the UH 
QUIRC camera (Hodapp et al.\ 1996) on the University of Hawaii 2.2 m 
telescope.  The total exposure 
time was 50 minutes, and we used standard IR reduction procedures.

\section{Results and Discussion}

\subsection{The BAL QSO}
The field of the QSO is shown in Fig.\ 1.  
The redshift of 
the QSO, determined from the \ion{He}{2} and \ion{C}{3}] broad emission lines,
is $z_{\rm QSO}$ = 2.169.  The coordinates, measured from the HST archival 
images, are $\alpha_{J2000} = 01^{\rm h}\ 37^{\rm m}\ 40\fs67$ and 
$\delta_{J2000} = 33\deg\ 08\arcmin\ 56\farcs8$.  From these same images,
we obtain $m_{F555W} = 21.14$ and $m_{F814W} = 21.28$.   Kirhakos et al.\ 
(1994) give $m_{g} = 21.2$.

Figure 2 shows the observed frame spectrum of the BAL QSO.  The only trough
evident in this region is \ion{C}{4}, which shows at least three 
components and outflow velocities
extending to $-$12000 km s$^{-1}$.   The trough is superposed on the emission
line, and the red edge is around +1,000 km s$^{-1}$, as can be seen clearly 
in Fig.\ 2. 

Unfortunately, \ion{Mg}{2} $\lambda 2800$ did not fall within our observed 
range, so we cannot determine whether this is a low-ionization
BAL QSO.  Weymann et al.\ (1991) found that the continua of low-ionization 
BAL QSOs are substantially redder than those of high-ionization BAL QSOs and 
non-BAL QSOs, presumably due to higher amounts of dust 
(Sprayberry \& Foltz 1992).
This BAL QSO appears reddened with respect to the Faint Object Spectrograph 
composite QSO spectrum (Zheng et al. 1997), but this reddening 
could be due to the intervening metal-line absorption system described in
\S3.2.   
There are hints of \ion{Fe}{2} and \ion{Fe}{3} emission
superposed on \ion{C}{3}] + \ion{Al}{3}, and a possible weak trough 
bluewards of this feature, both of which are common in low-ionization BAL QSOs.

\subsection{A Damped Ly$\alpha$ Absorber at $z_{\rm abs}=1.667$?}

Figure 2 shows several unresolved absorption features, which we
identify as arising from an intervening low-ionization metal-line
system at $z_{\rm abs}=1.667$.  Table 1 gives the line identifications
and measured equivalent widths.  The absorption features are unusually
strong, with Al III and Fe II rest-frame equivalent widths being of
strength similar to or greater than that seen in the relatively rare
damped Ly$\alpha$ systems (eg.\ Lu et al.\ 1996).  We have performed a
single-component curve-of-growth analysis on the five \ion{Fe}{2} lines
($b=30$ km s$^{-1}$), yielding an estimated \ion{Fe}{2} column density
of log $N$(\ion{Fe}{2})$=14.8$ cm$^{-2}$.  Assuming a solar metallicity
(log $N({\rm Fe})/N({\rm H})=-4.49$, Anders \& Grevesse 1989), this implies
the absorber has an \ion{H}{1} column density of
$N$(\ion{H}{1})$=2\times10^{19}$ cm$^{-2}$.  This is likely to be a
damped Ly$\alpha$ absorber since the metallicity of such systems is
typically $\approx10$\% solar (Pettini et al.\ 1997).  Furthermore, the
fact that Fe depletes readily onto dust in H I gas and our selection of
a conservatively high Doppler value in the curve-of-growth analysis
both add to the likelihood that this \ion{H}{1} column density estimate
is low by a factor of at least 10.

Low-ionization QSO metal-line absorption systems have been identified
with the gaseous halos of luminous galaxies (Bergeron \& Boiss\'{e}
1991; Steidel 1993) with large absorption cross sections, $D\approx
100$ kpc ($H_0=75$).  The damped absorbers, however, are believed
to arise from the inner disk regions (few 10s of kpc) of the same
absorbers based on their \ion{H}{1} column density (Wolfe et al.\ 1986), mass
density (Wolfe 1987), metallicity, and asymmetric absorption profile
kinematics (Prochaska \& Wolfe 1997).  Attempts to image high-redshift
damped Ly$\alpha$ absorbers have met with limited success, presumably
due in part to the close proximity of the intervening system to the QSO
image.  High-spatial-resolution {\it HST} images have recently revealed candidate
galaxies within a few arcsec of the QSO for several moderate-redshift
($0.4 < z < 1$) damped Ly$\alpha$ absorbers (Le Brun et al.\ 1997), but
the absorbers appear to be morphologically diverse.  The interpretation
of extended structure associated with this QSO is therefore likely to be
confused if indeed the intervening $z_{\rm abs}=1.667$ system arises
from a damped Ly$\alpha$ absorber lying very near the QSO line of
sight.

\subsection{Extended Luminous Material:  Intervening System or Host Galaxy?}

Figure \ref{fuzz} shows the extent and distribution of the faint, extended
luminous material around the QSO, from the HST archival images.  In both the 
F555W and F814W images, the overall impression is of a roughly elliptical
morphology, with a major axis at position angle $\sim25$\arcdeg, and a
center offset slightly to the SW from the QSO.  There is evidence for
a discrete object, or at least a luminosity peak, $\sim0\farcs7$ NE of
the QSO, present in both images but most obvious in the F814W image.
The southern edge of the ellipse appears
to form an arc-like feature (Fig.\ \ref{fuzz}$b$), but this impression
is dependent on only a few crucial pixels and could easily result
from noise fluctuations.

By interpolating over the central 0\farcs4 region of our PSF-subtracted
images (Fig.\ \ref{fuzz}$b$ and $d$), we can obtain a conservative
estimate of the magnitudes of the extended component alone.  We find
$m_{F555W}=23.8\pm0.1$ and $m_{F814W}=24.1\pm0.1$ in 2\farcs6-diameter
apertures, where the errors include
only measurement uncertainties.  If the luminosity were due to stars
at the redshift of the possible damped absorber ($z=1.667$), the slope of the
spectral-energy distribution at rest-frame 2000--3000 \AA, compared with
Bruzual \& Charlot (1997) models, indicates that it
would likely be dominated by a stellar population with an age of
$\sim2.5\times10^8$ years for solar metallicity, or roughly twice this
age for 20\% of solar metallicity.  On the other hand, if we are observing stars
in the QSO host galaxy, at $z=2.169$, the corresponding ages of the dominant 
population at rest-frame 1800--2500 \AA\ would be $\sim5\times10^8$ and
$\sim7.5\times10^8$ years, respectively.

We do not detect any extension in our deep, ground-based $H\!+\!K'$ image.
This image has a 1-$\sigma$ detection threshold of
$\sim5\times10^{-20}$ erg s$^{-1}$ cm$^{-2}$ \AA$^{-1}$ arcsec$^{-1}$.  For
a 2-$\sigma$ detection, our surface-brightness limit (in $f_{\lambda}$) is
essentially the same as that for the F814W filter, although we lose some
additional detection efficiency because we are looking in the wings of the
QSO profile.  In any case, our failure to detect the extension in the IR
is consistent with the stellar populations inferred above and would be
inconsistent with the presence of any additional red component that would
dominate either of these populations by a factor of $\sim5$ or more in the
observed IR (rest-frame 6000--7000 \AA).

Thus, we cannot distinguish between an intervening or host-galaxy origin
for the extended material in terms of our current knowledge of its SED.
A galaxy at $z=1.667$ dominated by a $\sim2.5\times10^8$-year-old population
would not be sufficiently unusual to constitute an objection to an intervening
origin.  We have been forced by our low S/N to deal with average colors, 
but inspection
of Fig.\ \ref{fuzz} indicates that the peak or discrete object just NE
of the QSO may be somewhat redder than the rest of the extended material.  It is
tempting to suggest that this object may be responsible for the intervening
system, and that the more diffuse extended material may be the QSO host
galaxy; but this suggestion is little more than speculation at this stage.

\acknowledgments

This research was partially supported by NSF under grant AST95-29078.  K.C.R.
acknowledges support provided by NASA through the Hubble Fellowship grant 
\#HF-01076.01-94A awarded by the Space Telescope Science Institute, which is 
operated by the Association of Universities for Research in Astronomy, Inc.,
for NASA under contract NAS 5-26555.

\newpage
 
\begin{center}
\begin{deluxetable}{lrrr}
\tablewidth{4.3in}
\tablecaption{Absorption Features in the $z_{\rm abs}=1.667$ System}
\tablehead{\multicolumn{1}{c}{$\lambda_{\rm obs}$} & \multicolumn{1}{c}{ID} & $W_\lambda^{\rm obs}$ (\AA)\tablenotemark{a}\hspace{-0.15in} & \hspace*{0.15in}$W_\lambda^{\rm rest}$ (\AA)}
\startdata
4947.9 & Al III 1854.72 & 1.13 & 0.42\hspace*{0.17in}\nl
4969.9 & Al III 1862.79 & 0.76 & 0.28\hspace*{0.17in}\nl
6252.9 & Fe II  2344.21 & 2.48 & 0.93\hspace*{0.17in}\nl
6333.7 & Fe II  2374.46 & 1.43 & 0.54\hspace*{0.17in}\nl
6355.2 & Fe II  2382.76 & 2.99 & 1.12\hspace*{0.17in}\nl
6899.2 & Fe II  2586.65 & 3.29 & 1.23\hspace*{0.17in}\nl
6935.3 & Fe II  2600.17 & 2.99 & 1.12\hspace*{0.17in}\nl
\enddata
\label{table:eqws}
\tablenotetext{a}{1$\sigma$ $W_\lambda$ error $\approx$ 120 m\AA\ observed, 45 m\AA\ in the rest frame.}
\end{deluxetable}
\end{center}

\begin{figure}
\caption{The field of 3C\,48 and the BAL QSO 0134+3253, from an HST WFPC2
F814W archival 
image.  The inset shows the BAL QSO at higher contrast and $3\times$ larger 
scale.  The parallel lines indicate the position and orientation of the 
spectroscopic slit.  North is up and East to the left.}

\caption{See next page.}

\caption{HST WFPC2 images of the extended material around the BAL QSO 0134+3253.(a) The F555W image, with a lower-contrast version in the inset.  (b)  The
F555W image, smoothed with a Gaussian with $\sigma=1$ pixel, and with a PSF
derived from a nearby star subtracted (see text for details).  (c) and (d)  Like(a) and (b), but for the F814W filter.}\label{fuzz}
\end{figure}

\setcounter{figure}{1}
\begin{figure}
\epsscale{1.0}
\plotone{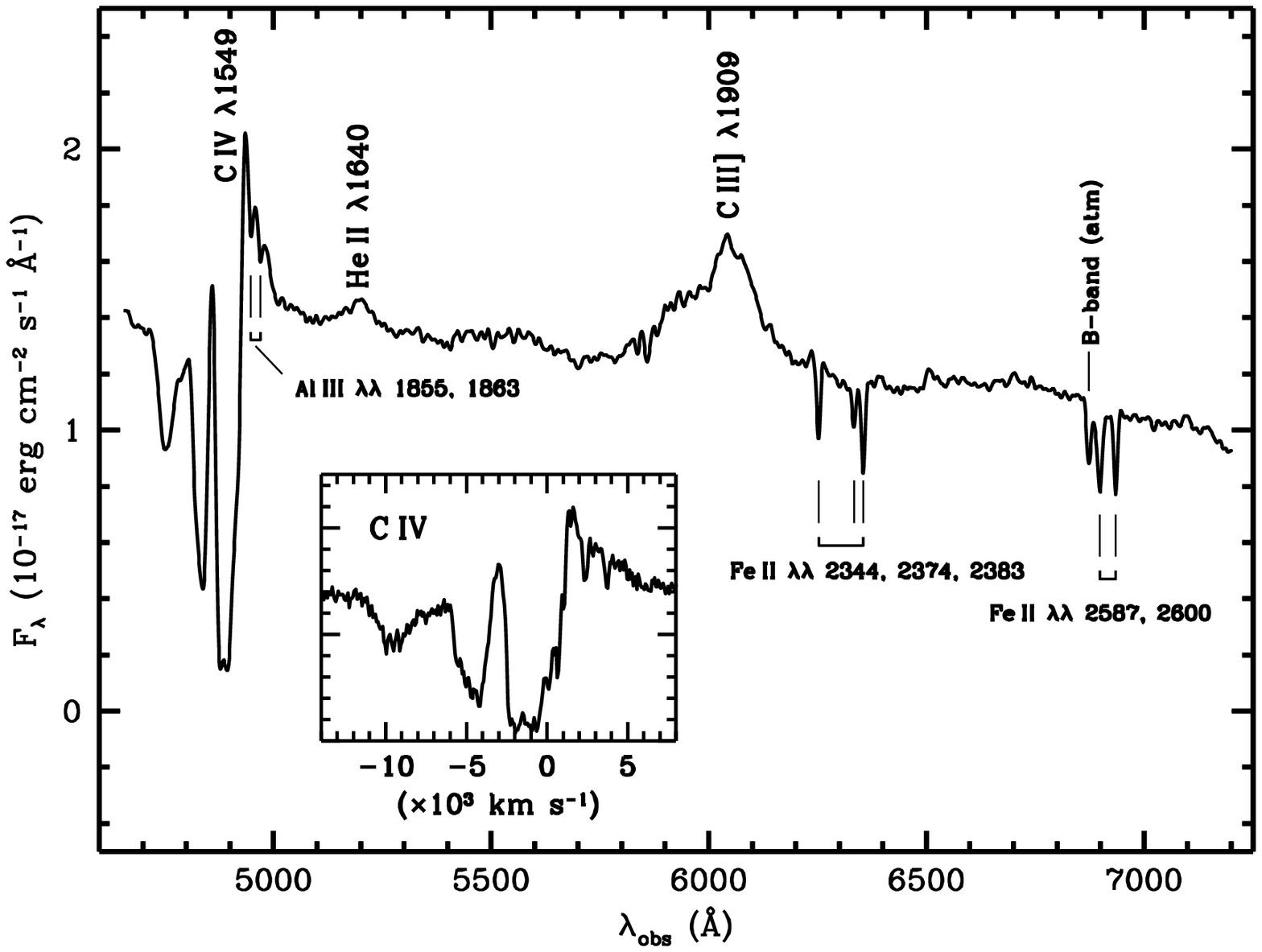}
\label{BALqso}
\caption{BAL QSO 0134+3253 spectrum in the observed frame.  The QSO redshift
from the broad emission lines is 2.169.  The spectrum has been 
smoothed with a Gaussian filter of $\sigma = 3$ \AA.  The narrow absorption
lines come from an intervening absorption system at  $z_{\rm abs}$ = 1.667.
The inset shows the C\,IV trough in the reference frame 
of the QSO.  The spectrum in the inset has not been smoothed.}
\end{figure}

\end{document}